# Triple Exponential Relaxation Dynamics in a Metallacrown-Based {Dy$^{III}$Cu$^{II}_5$} 3$d$-4$f$ Single-Molecule Magnet

Quan-Wen Li,[a] Rui-Chen Wan,[a] Jin Wang,[a] Yan-Cong Chen,[a] Jun-Liang Liu,[a]* Daniel Reta,[b] Nicholas F. Chilton,[b]* Zhen-Xing Wang,[c] and Ming-Liang Tong*[a]

[a]     Dr. Q.-W. Li, R.-C. Wan, J. Wang, Dr. Y.-C. Chen, Dr. J.-L. Liu, Prof. Dr. M.-L. Tong
Key Laboratory of Bioinorganic and Synthetic Chemistry of Ministry of Education, School of Chemistry, Sun Yat-Sen University, Guangzhou 510275 (P. R. China)
E-mail: liujliang5@mail.sysu.edu.cn (J.L.L.); tongml@mail.sysu.edu.cn (M.L.T.)
[b]     Dr. D. Reta, Dr. N. F. Chilton
School of Chemistry, The University of Manchester, Oxford Road, Manchester M13 9PL (UK)
E-mail: nicholas.chilton@manchester.ac.uk (N.F.C.)
[c]     Dr. Z.-X Wang
Wuhan National High Magnetic Center, Huazhong University of Science and Technology, Wuhan 430074 ( P. R. China)

**Abstract:** The interplay of strong single-ion anisotropy and magnetic interactions often give rise to novel magnetic behavior and can provide additional routes for controlling magnetization dynamics. However, novel effects arising from interactions between lanthanide and transition-metal ions are nowadays rarely observed. Herein, a {Dy$^{III}$Cu$^{II}_5$} 3$d$-4$f$ single-molecule magnet (SMM) is constructed as a rigid and planar [15-MC-5] metallacrown (MC), where the Dy$^{III}$ ion is trapped in the central pseudo-$D_{5h}$ pocket. A strong axial crystal field (CF) imbues the Dy$^{III}$ ion with large Ising-type magnetic anisotropy, and we are able to observe and model the magnetic interactions between the Cu$^{II}$-Cu$^{II}$ and Dy$^{III}$-Cu$^{II}$ pairs. Butterfly-shaped magnetic hysteresis shows clear steps at ±0.4 T, coincident with level crossings in our model exchange Hamiltonian between the {Cu$^{II}_5$} and Dy$^{III}$ spin systems. Most intriguingly, this air-stable SMM exhibits three distinct regimes in its magnetic relaxation dynamics, all clearly displaying an exponential dependence on temperature.

Single-molecule magnets (SMMs)[1,2] have attracted intensive attention as potential materials for ultra-high-density magnetic information storage and quantum information processing.[3-5] Recently there have been some significant breakthroughs on dysprosium single-ion magnets (SIMs), resulting in large effective energy barriers ($U_{eff}$) and high blocking temperatures ($T_B$).[6-16] There have also been many nice $d$-$f$ SMMs reported such as {M$^{II}_2$Dy$^{III}$} (M = Fe, Co)[17,18], {M$^{III}_2$Ln$^{III}_2$} (M = Cr, Co; Ln = Dy, *etc.*)[19,20], {U$^V_x$Mn$^{II}_y$} ($x$ = 12, $y$ = 6 or $x$ = 1, $y$ = 2)[21,22] and {Co$^{II}_2$Dy$^{III}_2$}[23], showing interesting magnetic dynamics such as exchange-biased magnetic hysteresis.

The main benefit of heterometallic $d$-$f$ SMMs over homometallic species is the ability to exploit hierarchical energy scales such as the crystal field (CF), and $d$-$d$/$d$-$f$ magnetic exchange interactions, which can introduce a higher density of magnetic states to provide additional routes for tuning magnetic relaxation pathways. It has become apparent in recent years that high-performance Dy$^{III}$ SIMs should be constructed with strong axial CF potentials and high-order symmetry axes (*e.g. $D_{4d}$, $D_{5h}$* and $C_{n>6}$), in order to suppress quantum tunneling of magnetization (QTM) by reducing mixing between opposing projections of the magnetic moment.[6] However, many of the recent developed species are coordinatively unsaturated and therefore usually highly reactive in air;[9,11-16] such inherent instability is a problem for progressing the field from academic research to industrial applicability, and therefore chemical stability under ambient conditions must also be a core goal for the community.[24] Therefore, we are keen to exploit a new strategy for designing 3$d$-4$f$ SMMs that are stable under ambient conditions by marrying high pseudo-symmetry at the 4$f$ ion with 3$d$-4$f$ magnetic exchange.

Considering the success of pentagonal bipyramidal lanthanide SIMs,[6-10] one strategy to achieve our goal is to place a 4$f$ ion into a pseudo-$D_{5h}$-symmetric pocket, with 3$d$ ions located at the periphery. This design typifies what is known as a metallacrown (MC), that can be assembled into various congeners to afford different geometries by tuning ditopic ligands and transition metal ions.[25] Inspired by this concept, our aim was to employ a [15-MC-5] congener to build a ~$D_{5h}$ symmetric pocket for the lanthanide ion, and to also foster magnetic exchange between the 3$d$ ions in the MC ring with the 4$f$ ion in the center. In addition to the high-order pseudo-symmetry axis, the axial ligands should be rich in electron density to provide a strong axial CF and stabilize the largest magnetic states of our target central ion, Dy$^{III}$.[26,27] Suitable ligands include phenoxides, phosphine oxides, alkoxides, carbanions and cyclopentadienyls, which have been proven to generate excellent SMMs.[6-16] Among these, phenoxide ligands stand out as being relatively stable in both air and polar solvents, which is beneficial for synthesizing, investigating and utilizing molecular materials under mild conditions.

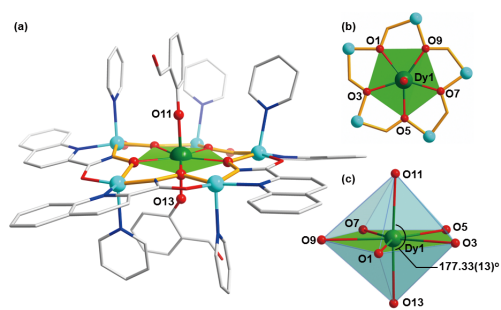

**Figure 1.** (a) Side view of the {Dy$^{III}$Cu$^{II}_5$} metallacrown; (b) Top view of the {Dy$^{III}$Cu$^{II}_5$} core; (c) Coordination environment of the Dy$^{III}$ ion. H atoms are omitted for clarity. Color codes: Dy, teal; Cu, pale blue; N, blue; O, red; C, light grey. The golden sticks show the [15-MC-5] metallacrown and the green pentagonal area show the planar coordination environment of Dy$^{III}$ ion.

We have utilized quinaldichydroxamic acid (H$_2$quinha) and Cu$^{II}$ to build a [15-MC-5] ring that encapsulates a single Dy$^{III}$ ion in the presence of salicylaldehyde (Hsal) and pyridine (see Supporting Information for synthetic details). Single-crystal X-ray diffraction reveals a hexanuclear {Dy$^{III}$Cu$^{II}_5$} complex, [DyCu$_5$(quinha)$_5$(sal)$_2$(py)$_5$][CF$_3$SO$_3$]·py·4H$_2$O (**1**), that crystallizes in the *P*-1 space group (Table S1) with the molecular structure shown in Figure 1. Five quinha$^{2-}$ ligands bridge five Cu$^{II}$ ions end to end, forming a [15-MC$_{quinha,Cu(II)}$-5] ring with a -[Cu-N-O]- repeat unit. The [15-MC-5] ring equatorially encapsulates a Dy$^{III}$ ion with five hydroximate oxygen donor atoms, while two phenoxide groups from the salicylaldehydes coordinate axially. The average equatorial and axial Dy-O distances are 2.413(3) and 2.198(3) Å, respectively. The equatorial O-Dy-O angles vary from 70.74(10)° to 70.33(10)° while the axial one is 177.33(13) (Table S2 and S3). This indicates that the Dy$^{III}$ possesses pentagonal bipyramidal geometry (Figure 1c), and is further confirmed by CShM analysis[28] with gives a small value of 0.244 for this geometry (Table S4). On the other hand, each Cu$^{II}$ ion on the MC ring is five-coordinate (square pyramidal, Table S5) with two quinha$^{2-}$ ligands providing two oxygen atoms and two nitrogen atoms to form the equatorial plane, and one axial pyridine. The adjacent intramolecular Cu···Cu and Cu···Dy distances are 4.595-4.643 Å and 3.904-3.964 Å, respectively. Nearest-neighbor {DyCu$_5$}$^+$ cores are well separated by triflate anions and solvent molecules, with the shortest intermolecular Dy···Dy distance being *ca.* 11 Å. To obtain complementary insight into the magnetic behavior of **1**, we synthesized its diamagnetic 4*f* analogue {Y$^{III}$Cu$^{II}_5$}, formulated as [YCu$_5$(quinha)$_5$(sal)$_2$(py)$_5$][CF$_3$SO$_3$]·py·4H$_2$O (**2**); all attempts to obtain the diamagnetic 3*d* analogue {Dy$^{III}$Zn$^{II}_5$} failed.

The magnetic susceptibilities of **1** and **2** were measured on polycrystalline samples under a 1 kOe direct-current (dc) field (Figure 2). Upon cooling, the $\chi_M T$ of **2** decreases from 1.58 cm$^3$ K mol$^{-1}$ at 300 K to 0.45 cm$^3$ K mol$^{-1}$ at 2 K, which is as expected for a spin-ground state of $S = 1/2$ (0.45 cm$^3$ K mol$^{-1}$, $g = 2.2$), indicating antiferromagnetic interactions between adjacent Cu$^{II}$ ions. For **1**, the $\chi_M T$ value at 300 K of 15.1 cm$^3$ K mol$^{-1}$ is close to the sum of one Curie-like Dy$^{III}$ ion (14.17 cm$^3$ K mol$^{-1}$, $^6H_{15/2}$, $g_J = 4/3$) and the room-temperature $\chi_M T$ value of **2** (1.58 cm$^3$ K mol$^{-1}$, sum of 15.75 cm$^3$ K mol$^{-1}$). On cooling, the $\chi_M T$ of **1** shows a gradual decline, owing to a combination of the CF splitting of the $^6H_{15/2}$ ground multiplet of Dy$^{III}$ and the antiferromagnetic interaction among Cu$^{II}$ ions. Below 15 K, there is a clear rise in $\chi_M T$, indicating the presence of ferromagnetic interactions between the Dy$^{III}$ and Cu$^{II}$ ions. At lower temperatures still, the $\chi_M T$ product drops; this could be due to magnetic blocking, which is suggested by the divergence of zero-field-cooled/field-cooled (ZFC/FC) magnetic susceptibilities below 6.5 K (Figure 2, inset), or to subtleties of the low-energy exchange manifold.

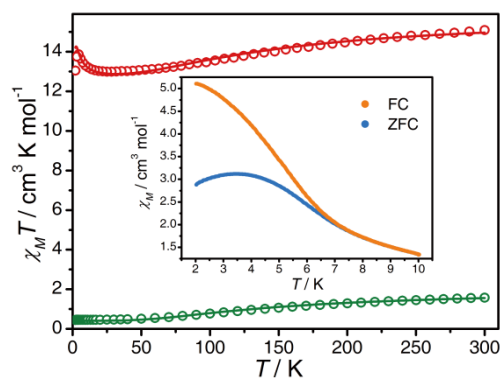

**Figure 2.** The $\chi_M T$ products measured under a 1 kOe dc field for **1** (red circles) and **2** (green circles). The solid lines are corresponding to the best fits described in the text. Inset: the ZFC(blue)/FC(orange) magnetic susceptibilities for **1** under a 2 kOe dc field sweeping at 2 K/min in warming mode.

To interpret the temperature dependence of the magnetic susceptibility of **1**, we first proceed to interpret that of the orbitally non-degenerate analogue **2**. As no zero-field splitting is possible for the local Cu$^{II}$ $S = 1/2$ states, the magnetic properties of **2** can be reliably described using the well-known Heisenberg-Dirac-van Vleck (HDVV) and Zeeman Hamiltonians (Equation 1). To minimize overparameterization, electron paramagnetic resonance (EPR) on a powder sample of **2** is employed to determine $g_{Cu\perp} = 2.03(2)$ and $g_{Cu\parallel} = 2.25(2)$ for the $S = 1/2$ ground state (Figure S13); given all Cu$^{II}$ sites have their anisotropy axes roughly parallel, we take these values also for the local Cu$^{II}$ $g$-values. The molecular structure of **2** suggests two types of Cu$^{II}$-Cu$^{II}$ magnetic interactions, those where neighboring Cu$^{II}$ ions have *syn-syn* or *syn-anti* arrangements of the coordinated pyridines (Figure 1 and Scheme 1); two approximate groups of magnetic interactions is also supported by DFT calculations (see ESI). Therefore, we fit the dc magnetic data of **2** with the *PHI* program[29] using a two-*J* model (Figure 2 and Scheme 1).

$$\hat{H} = -2J_{Cu-Cu}\left(\hat{S}_{Cu_1}\hat{S}_{Cu_2} + \hat{S}_{Cu_2}\hat{S}_{Cu_3} + \hat{S}_{Cu_4}\hat{S}_{Cu_5}\right) - 2J'_{Cu-Cu}\left(\hat{S}_{Cu_3}\hat{S}_{Cu_4} + \hat{S}_{Cu_1}\hat{S}_{Cu_5}\right) + \mu_B \bar{\bar{g}}_{Cu}\sum_{i=1}^{5}\left[\hat{S}_{Cu_i}\right]\vec{B} \quad (1)$$

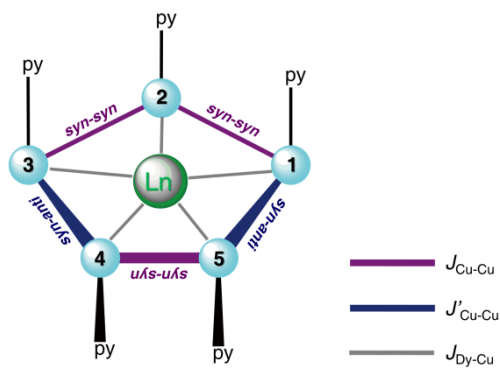

**Scheme 1.** The illustration of magnetic exchange couplings for {YCu$_5$} and {DyCu$_5$}.

The best-fit of the data gives $J_{Cu-Cu}$ = −44(2) cm$^{-1}$ and $J'_{Cu-Cu}$ = −68(2) cm$^{-1}$ with $g_{Cu\perp}$ and $g_{Cu\parallel}$ fixed from EPR; the ratio of the fitted exchange values is in good agreement with that extracted from broken-symmetry DFT calculations (Table S6 and S8). This results in an $S = 1/2$ ground state with a first excited $S = 1/2$ state at 51 cm$^{-1}$ (Table S9).

To proceed to modelling the dc magnetic data for **1**, we require a model for the CF states for the Dy$^{III}$ ion. Complete active space self-consistent field spin-orbit (CASSCF-SO) calculations (see ESI for details) on the Dy$^{III}$ site of **1** reveal that the ground and first excited Kramers doublets are well described by the $|\pm 15/2\rangle$ and $|\pm 13/2\rangle$ states, respectively, quantized along approximately the O11-O13 axis (Figure 1), separated by 314 cm$^{-1}$ (Table S10). We model the Dy$^{III}$-Cu$^{II}$ interactions with a Lines-type exchange interaction between the true spin of Dy$^{III}$, $S_{Dy} = 5/2$, and the $S_{Cu} = 1/2$ Cu$^{II}$ ions;[30] the CF parameters for Dy$^{III}$ are fixed from CASSCF-SO (Table S11), along with the Cu$^{II}$ parameters from the fitted data for **2**, and thus there are only two fitting parameters viz. $g_J$ and $J_{Dy-Cu}$ (Equation 2). The experimental dc magnetic data are well described with $J_{Dy-Cu}$ = +0.88(5) cm$^{-1}$ and $g_J$ = 1.30(1) (Figures 2 and S3), the latter being only slightly lower than the free-ion value of $4/3$, presumably accounting for the formation of molecular orbitals,[31] in contrast to the pure free-ion basis functions used in *PHI* (indeed, CASSCF-SO predicts $g_J$ = 1.32).

$$\hat{H} = -2J_{Cu-Cu}(\hat{S}_{Cu_1}\hat{S}_{Cu_2} + \hat{S}_{Cu_2}\hat{S}_{Cu_3} + \hat{S}_{Cu_4}\hat{S}_{Cu_5}) - 2J'_{Cu-Cu}(\hat{S}_{Cu_3}\hat{S}_{Cu_4} + \hat{S}_{Cu_1}\hat{S}_{Cu_5}) - 2J_{Dy-Cu}\hat{S}_{Dy}\sum_{i=1}^{5}[\hat{S}_{Cu_i}] + \mu_B(\bar{\bar{g}}_{Cu}\sum_{i=1}^{5}[\hat{S}_{Cu_i}] + g_J\hat{J})\vec{B} + \sum_{k=2,4,6}\sum_{q=-k}^{k} B_k^q \hat{O}_k^q \quad (2)$$

As the exchange interaction between Dy$^{III}$ and Cu$^{II}$ is so much smaller than the Dy$^{III}$ CF, $J_{Cu-Cu}$ and $J'_{Cu-Cu}$ parameters, the low-lying electronic states for **1** are well described as simple product functions between the Dy$^{III}$ and {Cu$^{II}_5$} spin systems (Table S12). Although the overall molecule is non-Kramers, many of the resulting states are pseudo-doublets; the low-lying states are sequential excitations of the {Cu$^{II}_5$} spin system (Table S9) coupled to the ground $|\pm 15/2\rangle$ state of Dy$^{III}$. In zero dc field, the low-lying pseudo-doublets arising from the Dy$^{III}$−Cu$^{II}$ exchange interactions are linear combinations of the total angular momentum along the z axis $M_z$. Upon application of a small magnetic field along the z-axis, the degeneracy of the pseudo-doublets is lost and the linear combinations are resolved into components of $\pm M_z$ (Figure 3).

The dc magnetization data for **1** exhibit butterfly-shaped hysteresis loops at low temperatures (Figure 4) that remain open up to 12 K at a sweep rate of 0.02 T/s. The hysteresis loops are closed at zero-field due to fast relaxation from the superimposed $M_z$ states, but rapidly open upon application of the field. There is an obvious step in both sweep directions at around 0.4 T (Figure 4b), consistent with a level crossing predicted at *ca.* 0.48 T from our exchange model (Figure 4c) between the ground and the first excited doublet arising from the Dy$^{III}$−Cu$^{II}$ exchange bias; this provides further validation of our exchange model.

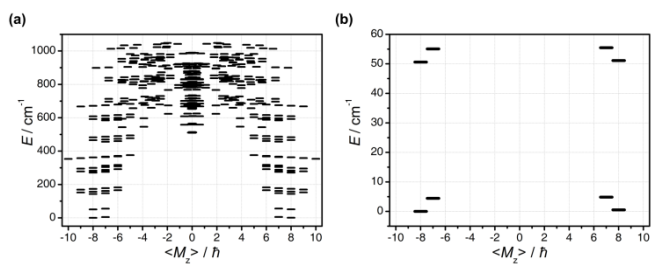

**Figure 3.** Simulated energy diagram of {DyCu$_5$}, showing the expectation value of the total angular momentum projection onto the z-axis (<$M_z$>) for each state. Simulations use the CF parameters obtained from *ab initio* calculations and other parameters extracted from fitting the magnetic and EPR data, as employed to model that magnetic susceptibility in Figure 2, using Equation 2. The external dc field applied along the z-axis is 0.5 kOe.

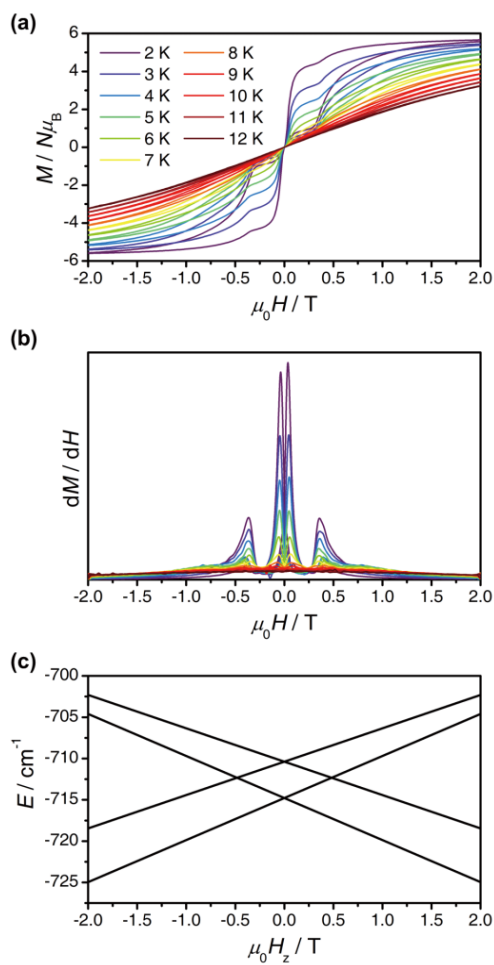

**Figure 4.** (a) Magnetic hysteresis loops for **1**. The data were continuously collected at time intervals of 1 s with the field sweeping rate of 0.02 T/s at various temperatures. (b) 1$^{st}$ derivative of magnetization vs. applied field. (c) Zeeman splitting of the four lowest magnetic states along the *z*-axis, calculated using Equation 2.

Given that **1** shows magnetic hysteresis, we have studied its magnetic relaxation properties. Alternating-current (ac) measurements show clear temperature- and frequency-dependent out-of-phase signals (Figure S4-6), from which we extract the relaxation times ($\tau$) for **1** by fitting with the generalized Debye model (Figure S7). Under zero dc field, an Arrhenius plot reveals two linear regions (Figure 6, red circles in green and red zones), characteristic of exponential temperature dependencies. [Analogous ac measurements for the yttrium analogue **2** do not show any signs of slow magnetic relaxation under zero field, revealing that this behavior cannot owe to the {Cu$^{II}_5$} ring alone.] Applying an optimized field of 2 kOe (Figure S6), the resulting relaxation times also exhibit two linear regions (Figure 6, blue circles in green and blue zones). Both datasets in the high temperature region (64–45 K) are nearly identical, while those in the low temperature region (<45 K) are very different, revealing three different slopes in all. We note that these relaxation profiles do not show power-law dependencies (as they all curve on a log-log plot of $\tau^{-1}$ vs. *T*, Figure S9) and thus cannot be ascribed to Raman-like processes (Figure S10).

Given the large change in the relaxation profile under an applied dc field, we were curious about the relaxation dynamics under an intermediate dc field between 0 and 2 kOe. With an applied dc field of 500 Oe, we curiously observe three unique domains in the relaxation dynamics (Figures 5 and S8), which also all show exponential temperature dependencies (Figure 6, green circles). At high temperatures, the relaxation times are nearly identical to the 0 and 2 kOe data (green zone), and below 45 K the slope becomes smaller and appears very similar to that under 2 kOe (7–45 K, blue zone). On further lowering the temperature the slope decreases again (2.5–6 K, red zone), and is parallel to that observed under zero dc field (2.5–45 K, red zone). Similarly to the 0 and 2 kOe data, these three domains are linear on an Arrhenius plot and are curved on a log-log plot of $\tau^{-1}$ vs. *T* (Figures 6 and S9) and thus clearly have an exponential, and not a power-law, dependence on temperature.

After observing three distinct domains in the 500 Oe relaxation data, we were curious if the dynamics under a 2 kOe field would also show a 3$^{rd}$ domain at lower temperatures. Indeed, dc relaxation experiments at 2 kOe (Figure S12) yield relaxation times that show a low slope between 2.5–5 K (Figure 6, blue diamonds in red zone), similar to the zero dc field (2.5–45 K, red zone) and 500 Oe (2.5–6 K, red zone) data. The presence of three unique exponential processes contrasts to the more common observation of either one or two processes combined with other relaxation mechanisms,[6-12] or the different case of multiple relaxation times under the same conditions.[13,16,32-35]

All three datasets can be fitted simultaneously with a sum of three exponential terms (Equation 3), with three common exponents ($U_{eff}$) and field-dependent pre-exponential factors ($\tau_0$). We note pre-emptively that the phenomenological $U_{eff}$ are not necessarily attributed to traditional Orbach process. Fitting the data gives $U_{eff(1)}$ = 623(22) cm$^{-1}$ (896(31) K), $U_{eff(2)}$ = 38.7(5) cm$^{-1}$ (55.7(7) K), and $U_{eff(3)}$ = 5.2(2) cm$^{-1}$ (7.5(3) K), with $\tau_0$ values in Table 1.

$$\tau^{-1} = \tau_{0(1)}^{-1}\exp(-U_{eff(1)}/k_BT) + \tau_{0(2)}^{-1}\exp(-U_{eff(2)}/k_BT) + \tau_{0(3)}^{-1}\exp(-U_{eff(3)}/k_BT) \quad (3)$$

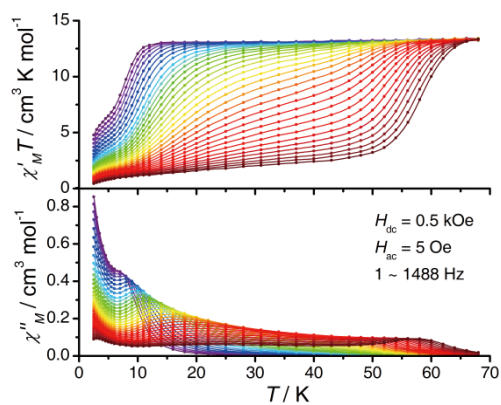

**Figure 5.** The temperature-dependent ac susceptibilities for **1** under a 0.5 kOe dc field with the frequency of 1−1488 Hz. The solid lines are guide for eyes.

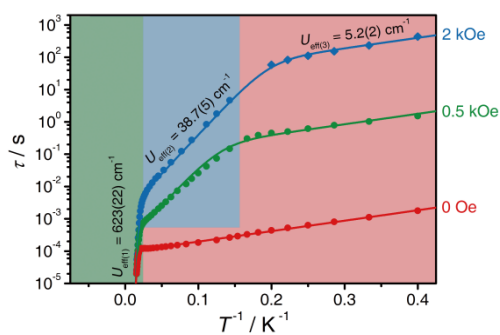

**Figure 6.** The temperature-dependent relaxation times for **1** under zero dc field (red circles), 0.5 kOe dc field (green circles) and 2 kOe dc field (blue circles from ac magnetic susceptibility and blue diamond from dc magnetization decay as shown in Figure S12). The blue, red and green lines are Arrhenius law fits for 0, 0.5 and 2 kOe dc field (high [green], intermediate [blue] and low [red] temperature region), respectively.

**Table 1.** Effective energy barriers ($U_{eff}$) and pre-exponential factors ($\tau_0$) for complex **1**.

|  | 0 Oe | 0.5 kOe | 2 kOe |
| --- | --- | --- | --- |
| $U_{eff(1)}$ [cm$^{-1}$] |  | 623(22) |  |
| $U_{eff(2)}$ [cm$^{-1}$] | - |  | 38.7(5) |
| $U_{eff(3)}$ [cm$^{-1}$] |  | 5.2(2) |  |
| $\tau_{0(1)}$ [s] | 2.1(11)×10$^{-11}$ | 2.4(12)×10$^{-11}$ | 2.7(15)×10$^{-11}$ |
| $\tau_{0(2)}$ [s] | - | 1.63(8)×10$^{-4}$ | 1.53(9)×10$^{-3}$ |
| $\tau_{0(3)}$ [s] | 9.4(6)×10$^{-5}$ | 9.0(8)×10$^{-2}$ | 2.0(2)×10$^{1}$ |

The best-fit parameters (Table 1) show negligible field dependence for the pre-exponential factors of the 1$^{st}$ term, $\tau_{0(1)}$, as expected given the data are practically coincident. However, we observe a marked field dependence for the other two, $\tau_{0(2)}$ and $\tau_{0(3)}$. The large exponent ($U_{eff(1)}$ = 623(22) cm$^{-1}$) for the 1$^{st}$ process is believed to be due to a true Orbach relaxation process involving the CF states of the Dy$^{III}$ ion, and relaxation likely occurs via a set of highly mixed states at 510−611 cm$^{-1}$ (Figure 3a). To the best of our knowledge, $U_{eff(1)}$ = 623 cm$^{-1}$ is the largest energy barrier for any reported d-f SMM, eclipsing $U_{eff}$ = 416 cm$^{-1}$ for a {Co$^{II}_2$Dy$^{III}$} complex.[18] For the 2$^{nd}$ and 3$^{rd}$ relaxation processes we note that $U_{eff(3)}$ = 5.2 and $U_{eff(2)}$ = 38.7 cm$^{-1}$ are similar to the 1$^{st}$ and 2$^{nd}$ excited states determined from our exchange model of **1** (4.4 and 50.6 cm$^{-1}$, respectively; Figure 3b and Table S12), and therefore these relaxation mechanisms could conceivably involve combined Dy$^{III}$ and Cu$^{II}$ spin-flip processes. However, we do not endorse such an assertion without reservation, and undoubtedly a thorough temperature- and field-dependent study is necessary to determine the nature of these two relaxation processes.

By utilizing a [15-MC-5] metallacrown and salicylaldehyde ligand, we have prepared a unique 3*d*-4*f* SMM with ferromagnetic 3*d*-4*f* interactions. In conjunction with DFT and CASSCF-SO calculations, we have modelled the static magnetic properties of the molecule, providing a plausible origin for the steps in the magnetic hysteresis at 0.4 T. Intriguingly, we observe three different exponential relaxation processes, the largest of which has $U_{eff(1)}$ = 623(22) cm$^{-1}$ (a record for *d-f* SMMs), that likely arises from the Dy$^{III}$ single-ion anisotropy. The other two exponential processes have strong field dependence, and as yet we do not understand their origins. However, we suggest that the interplay between the hierarchical energy scales of the Dy$^{III}$ anisotropy, Cu$^{II}$−Cu$^{II}$ exchange and Dy$^{III}$−Cu$^{II}$ exchange are central to these dynamics. Unraveling the nature, origins and implications of these three distinct relaxation domains will be the subject of an upcoming full study.

**Acknowledgements**

This work was supported by the "973 Project" (2014CB845602), the NSFC (Grant nos 21620102002, 21773316, 21371183 and 91422302, 21701198), the Fundamental Research Funds for the Central Universities (Grant 17lgjc13 and 17lgpy81), China Postdoctoral Science Foundation (National Postdoctoral Program for Innovative Talents, Grant BX201700295), the Ramsay Memorial Trust (fellowship to NFC), the EPSRC (EP/P002560/1) and The University of Manchester.

# Supporting Information

**Table of Contents**

## S1. Experimental Section

**Materials and Physical Measurements.** The ligand $H_2$quinHA was synthesized as literature described.[1] Metal salts and other reagents were commercially available and used as received without further purification. The C, H, and N elemental analyses were carried out with an Elementar Vario-EL CHNS elemental analyzer. Powder X-ray diffraction (PXRD) patterns were performed on Bruker D8 Advance Diffractometer (Cu-$K_\alpha$, $\lambda$ = 1.54056 Å). Thermogravimetric analysis (TGA) was carried out on a NETZSCH TG209F3 thermogravimetric analyzer. Magnetic susceptibility measurements were performed with a Quantum Design MPMS-XL7 SQUID. The ZFC-FC (0.2 T, 2 K/min), hysteresis (0.02 T/s) were measured on a Quantum Design PPMS. Polycrystalline samples were embedded in vaseline to prevent torqueing. Data were corrected for the diamagnetic contribution calculated from Pascal constants.

**Single Crystal X-ray Crystallography.** Diffraction data were collected on a Bruker D8 QUEST diffractometer with Mo-$K_\alpha$ radiation ($\lambda$ = 0.71073 Å) for complexes **1** and **2** at 120(2) K. The Data indexing and integration were carried out using a Bruker Smart program. The structures were solved by direct methods, and all non-hydrogen atoms were refined anisotropically by least-squares on $F^2$ using the SHELXTL program suite.[2] Anisotropic thermal parameters were assigned to all non-hydrogen atoms. The hydrogen atoms attached to carbon, nitrogen and oxygen atoms were placed in idealised positions and refined using a riding model to the atom to which they were attached. The SQUEEZE program of PLATON was employed to deal with the disordered solvent molecules.[3] The solvent accessible volume and the number of residual electrons per unit cell is 330 Å$^3$, 89e and 342 Å$^3$, 87e for **1** and **2**, respectively. The elemental analysis and thermogravimetric analysis were applied to determine the disordered solvent molecules as three water molecules, which are consistent with the analysis results of SQUEEZE. CCDC 1563751 (**1**) and 1563752 (**2**) contain the supplementary crystallographic data for this paper. These data can be obtained free of charge via https://www.ccdc.cam.ac.uk/structures/.

**Synthesis.** [DyCu$_5$(quinha)$_5$(sal)$_2$(py)$_5$](CF$_3$SO$_3$)·py·4H$_2$O (**1**): A mixture of H$_2$quinHA (0.125 mmol, 23.5 mg) and Cu(CF$_3$SO$_3$)$_2$ (0.125 mmol, 45.2 mg) were dissolved in 5 mL MeOH. Under stirring, Dy(CF$_3$SO$_3$)$_3$ (0.025 mmol, 15.2 mg) was added into the solution. Then solution was transferred to a Teflon container and kept at 75 °C in the oven for 12 hr. Green microcrystals (Yield: 30 mg) were collected as precursor for the next steps. The precursor (20 mg) was dissolved in 5 mL MeOH and 1 mL pyridine. Under stirring, salicylaldehyde (Hsal, 0.2 mmol, 24.4 mg) was added. Then, the solution was stirred for 2 hr. After that, the solution was filtered and kept silent for evaporation. Green crystals (Yield, 15 mg, 38 % based on Dy) suitable for X-ray analysis were obtained in two weeks. Elemental analysis calcd (%) for DyCu$_5$C$_{95}$H$_{78}$N$_{16}$O$_{21}$SF$_3$; C: 48.57, H: 3.35, N: 9.54; found (%): C: 48.68, H: 3.57, N: 9.36.

[YCu$_5$(quinha)$_5$(sal)$_2$(py)$_5$](CF$_3$SO$_3$)·py·4H$_2$O (**2**): the procedure was the same as that employed for complex **1**, except that Dy(CF$_3$SO$_3$)$_3$ was replaced by Y(CF$_3$SO$_3$)$_3$ (Yield 13.4 mg, 34% based on Y). Green crystals were obtained in two weeks by slow evaporation of the solution. Elemental analysis calcd (%) for YCu$_5$C$_{95}$H$_{78}$N$_{16}$O$_{21}$SF$_3$; C: 50.15, H: 3.46, N: 9.85; found (%): C: 49.98, H: 3.61, N: 9.72.

## S2. Crystal Data and Structures

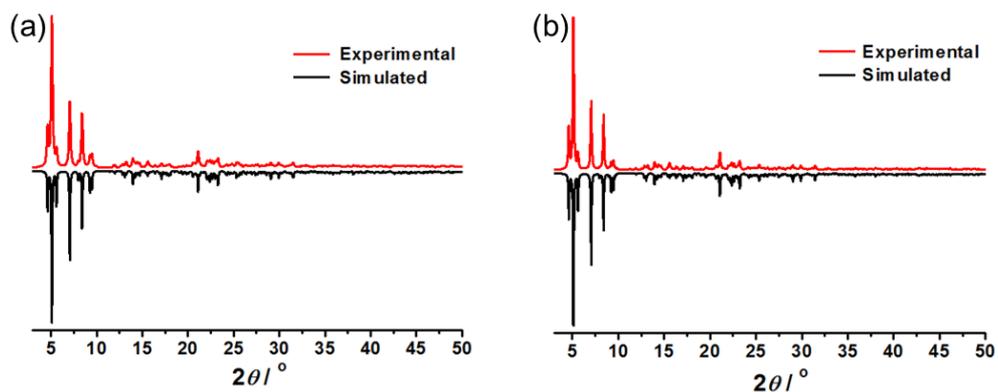

**Figure S1.** Experimental and simulated X-ray powder diffraction (XRPD) patterns for **1** (a) and **2** (b).

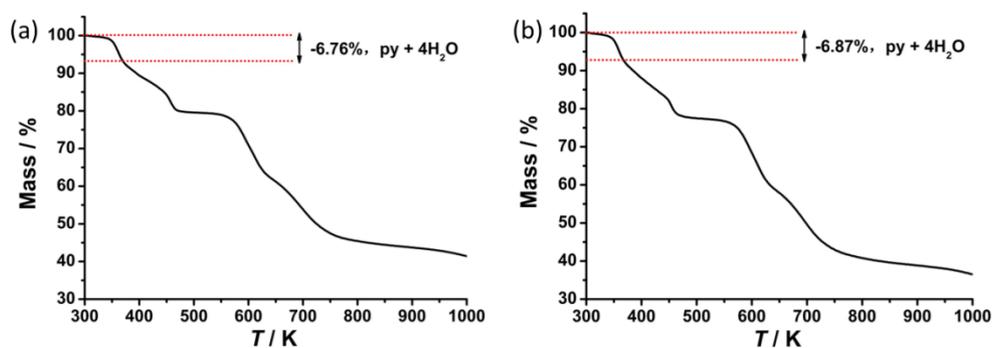

**Figure S2.** Thermogravimetric analysis of **1** (a) and **2** (b). The red dash lines show the stage of escaping of solvent molecules.

**Table S1.** Crystallographic data and structural refinements for **1** and **2**.

| Compound | 1 | 2 |
|---|---|---|
| Formula | $DyCu_5C_{95}H_{78}N_{16}O_{21}SF_3$ | $YCu_5C_{95}H_{78}N_{16}O_{21}SF_3$ |
| Formula weight | 2348.99 | 2275.40 |
| Temperature / K | 120(2) | 120(2) |
| Crystal system | Triclinic | Triclinic |
| Space group | *P*-1 | *P*-1 |
| a / Å | 13.3601(5) | 13.3721(5) |
| b / Å | 18.4387(8) | 18.3699(7) |
| c / Å | 21.1192(8) | 21.2229(8) |
| α / ° | 108.139(1) | 107.9971(13) |
| β / ° | 106.261(1) | 106.4516(13) |
| γ / ° | 92.988(1) | 92.7421(13) |
| V / Å$^3$ | 4690.6(3) | 4702.6(3) |
| Z | 2 | 2 |
| $\rho_{calcd.}$ (g/cm$^3$) | 1.663 | 1.607 |
| µ (mm$^{-1}$) | 2.009 | 1.830 |
| F (000) | 2364 | 2310 |
| Reflections collected | 67814 | 81364 |
| Independent reflections | 21351 | 21536 |
| GOF on $F^2$ | 1.052 | 1.027 |
| $R_1$, $wR_2$ [$I ≥ 2σ(I)$, squeeze]$^a$ | 0.0486, 0.1122 | 0.0544, 0.1448 |
| $R_1$, $wR_2$ (all data, squeeze) | 0.0724, 0.1247 | 0.0739, 0.1584 |
| CCDC No. | 1563751 | 1563752 |

$^a$ $R_1 = \sum||Fo| - |Fc||/\sum|Fo|$. $wR_2 = [\sum w(Fo^2 - Fc^2)^2/\sum w(Fo^2)^2]^{1/2}$.

**Table S2.** Selected bonds lengths [Å] and angles [°] for **1** and **2**.

| 1 | | 2 | |
|---|---|---|---|
| Dy1-O1 | 2.449(3) | Y1-O1 | 2.441(3) |
| Dy1-O3 | 2.408(3) | Y1-O3 | 2.398(3) |
| Dy1-O5 | 2.405(3) | Y1-O5 | 2.390(3) |
| Dy1-O7 | 2.408(3) | Y1-O7 | 2.392(2) |
| Dy1-O9 | 2.394(3) | Y1-O9 | 2.382(3) |
| Dy1-O11 | 2.197(3) | Y1-O11 | 2.195(3) |
| Dy1-O13 | 2.199(3) | Y1-O13 | 2.196(3) |
| O1-Dy1-O3 | 70.74(10) | O1-Y1-O3 | 70.83(9) |
| O3-Dy1-O5 | 73.33(10) | O3-Y1-O5 | 73.25(9) |
| O5-Dy1-O7 | 71.78(10) | O5-Y1-O7 | 71.80(9) |
| O7-Dy1-O9 | 72.21(10) | O7-Y1-O9 | 72.27(9) |
| O9-Dy1-O1 | 72.06(10) | O9-Y1-O1 | 71.97(9) |
| O11-Dy1-O13 | 177.33(13) | O11-Y1-O13 | 177.48(12) |

**Table S3.** Comparison of equatorial and axial Dy-O bonds in pentagonal bipyramidal Dy-based SMMs.

| Compounds | Dy–O$_{equatorial}$ bonds (Å) | Dy–O$_{axial}$ bonds (Å) | O$_{axial}$–Dy–O$_{axial}$ angle (°) | Ref |
|---|---|---|---|---|
| **1** | 2.394-2.449 | 2.197, 2.199 | 177.33 | This work |
| [Zn$_2$Dy(L$^1$)$_2$(MeOH)]NO$_3$·3MeOH·H$_2$O | 2.366-2.427 | 2.195, 2.221 | 168.6 | 1 |
| [Fe$_2$Dy(L$^2$)$_2$(H$_2$O)]ClO$_4$·2H$_2$O | 2.324-2.492 | 2.190, 2.193 | 169.2 | 2 |
| [Co$_2$Dy(L$^1$)$_2$(H$_2$O)]NO$_3$·3H$_2$O | 2.355-2.420 | 2.175, 2.198 | 169.7 | 3 |
| [Co$_2$Dy(L$^1$)$_2$(H$_2$O)]NO$_3$ | 2.326-2.508 | 2.171, 2.175 | 169.8 | 3 |
| [Dy(Cy$_3$PO)$_2$(H$_2$O)$_5$]Cl$_3$·(Cy$_3$PO)·H$_2$O·EtOH | 2.327-2.380 | 2.217, 2.221 | 175.79 | 4 |
| [Dy(Cy$_3$PO)$_2$(H$_2$O)$_5$]Br$_3$·2(Cy$_3$PO)·2H$_2$O·2EtOH | 2.336-2.365 | 2.189, 2.210 | 179.04 | 4 |
| [Dy(L$^3$)$_2$(H$_2$O)$_5$]I$_3$·(L$^3$)$_2$·H$_2$O | 2.355-2.375 | 2.203, 2.208 | 175.14 | 5 |
| [Dy(bbpen)Cl] | / | 2.166 | 154.3 | 6 |
| [Dy(bbpen)Br] | / | 2.163 | 155.8 | 6 |
| [Dy(O$^t$Bu)$_2$(py)$_5$](BPh$_4$) | / | 2.110, 2.114 | 178.91 | 7 |

* L$^1$ = 2,2′,2″-(((nitrilotris(ethane-2,1-diyl))tris(azanediyl))tris(methylene))tris-(4-bromophenol);
L$^2$ = 2,2′,2″-(((nitrilotris(ethane-2,1-diyl))tris(azanediyl))tris(methylene))tris-(4-chlorophenol));
L$^3$ = $^t$BuPO(NH$^i$Pr)$_2$;
H$_2$bbpen = N,N′-bis(2-hydroxybenzyl)-N,N′-bis(2-methylpyridyl)ethylenediamine.

**Table S4.** Continuous shape measures calculations (CShM) for rare-earth ions in **1** and **2**.

| Complex | HP-7 ($D_{7h}$) | HPY-7 ($C_{6v}$) | PBPY-7 ($D_{5h}$) | COC-7 ($C_{3v}$) | CTPR-7 ($C_{2v}$) | JPBPY-7 ($D_{5h}$) | JETPY-7 ($C_{3v}$) |
|---|---|---|---|---|---|---|---|
| Dy in **1** | 33.352 | 24.487 | 0.244 | 7.845 | 5.996 | 2.304 | 23.844 |
| Y in **2** | 33.316 | 24.555 | 0.233 | 7.826 | 5.999 | 2.357 | 23.885 |

HP-7 = Heptagon; HPY-7 = Hexagonal pyramid; PBPY-7 = Pentagonal bipyramid; COC-7 = Capped octahedron; CTPR-7 = Capped trigonal prism; JPBPY-7 = Johnson pentagonal bipyramid J13; JETPY-7 = Johnson elongated triangular pyramid J7.

**Table S5.** Continuous Shape Measures (CShM) calculations for Cu$^{II}$ ions in **1** and **2**.

|  | PP-5 | vOC-5 | TBPY-5 | SPY-5 | JTBPY-5 |
|---|---|---|---|---|---|
| complex **1** | | | | | |
| Cu1 | 25.838 | 2.174 | 5.377 | 1.350 | 8.893 |
| Cu2 | 24.102 | 2.395 | 5.959 | 1.705 | 8.349 |
| Cu3 | 28.548 | 2.005 | 4.214 | 0.886 | 6.767 |
| Cu4 | 26.464 | 2.048 | 5.410 | 1.226 | 8.930 |
| Cu5 | 26.113 | 1.821 | 6.003 | 1.307 | 8.242 |
| complex **2** | | | | | |
| Cu1 | 25.676 | 2.265 | 5.413 | 1.387 | 9.026 |
| Cu2 | 24.420 | 2.339 | 5.868 | 1.630 | 8.248 |
| Cu3 | 28.686 | 2.001 | 4.218 | 0.859 | 6.764 |
| Cu4 | 26.424 | 2.070 | 5.378 | 1.231 | 8.970 |
| Cu5 | 26.065 | 1.828 | 5.898 | 1.299 | 8.127 |

*PP-5 = Pentagon; vOC-5 = Vacant octahedron; TBPY-5 = Trigonal bipyramid; SPY-5 = Spherical square pyramid; JTBPY-5 = Johnson Trigonal.

## S3. Magnetic Characterization

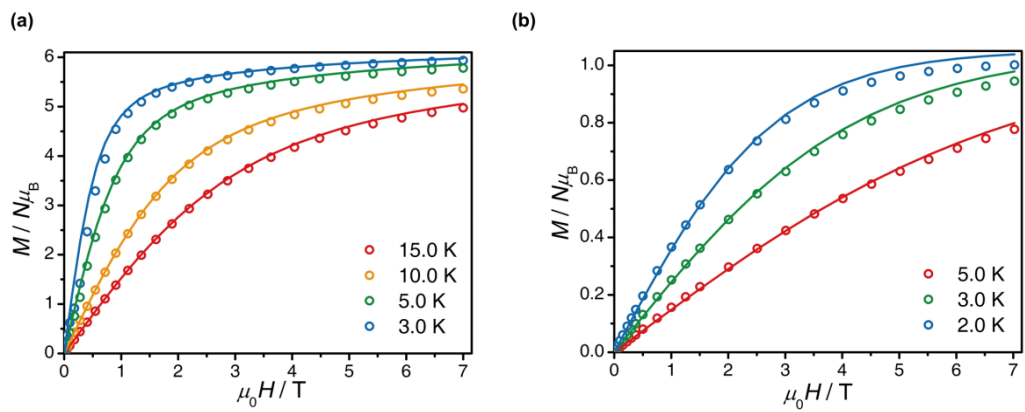

**Figure S3.** Variable-field magnetization data for **1** (a) and **2** (b). The solid lines are simulation from the fitted parameters for **1** and the best fit for **2** by using *PHI* program. Data were collected from 0−7 T in steady fields.

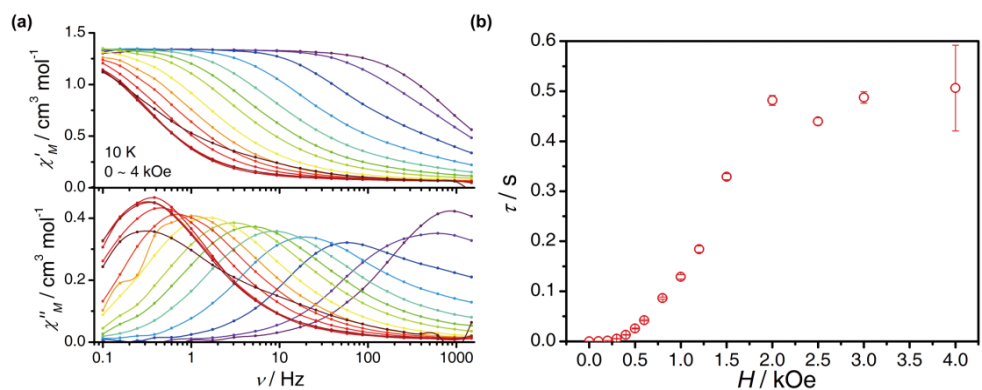

**Figure S4.** (a) Frequency dependence of the in-phase product ($\chi'_M$) and out-of-phase ($\chi''_M$) at 10 K under different applied fields (0−4 kOe) for **1**. The solid lines are guides for the eyes. (b) Field-dependent relaxation times for **1** at 10 K.

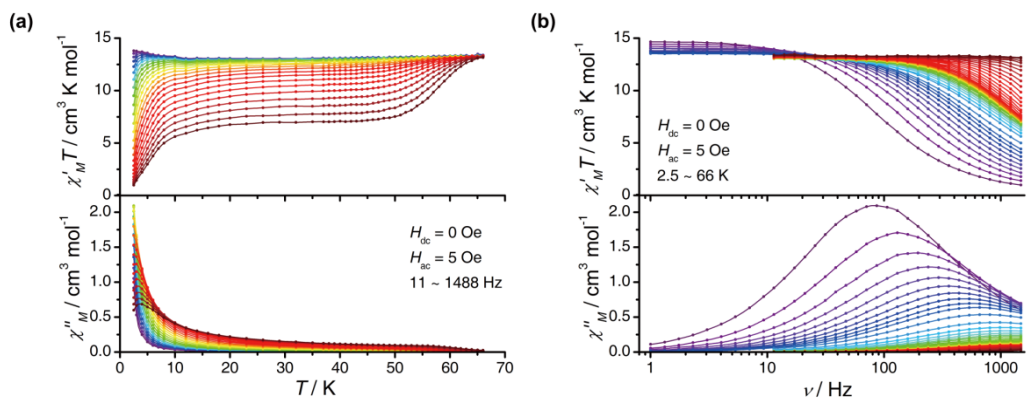

**Figure S5.** (a) The temperature dependence of the in-phase product ($\chi'_M T$) and out-of-phase ($\chi''_M$) for **1** at zero dc field with the ac frequency of 11−1488 Hz. (b) The frequency dependence of the in-phase product ($\chi'_M T$) and out-of-phase ($\chi''_M$) for **1** at zero dc field with the temperature of 2.5−66 K. The solid lines are guides for the eyes.

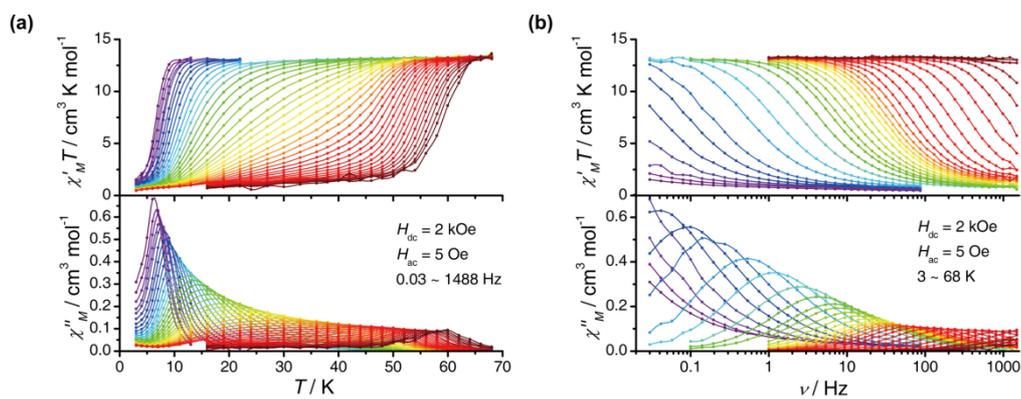

**Figure S6.** (a) The temperature dependence of the in-phase product ($\chi'_M T$) and out-of-phase ($\chi''_M$) for **1** at 2 kOe dc field with the ac frequency of 0.03−1488 Hz. (b) The frequency dependence of the in-phase product ($\chi'_M T$) and out-of-phase ($\chi''_M$) for **1** at 2 kOe dc field with the temperature of 3−68 K. The solid lines are guides for the eyes.

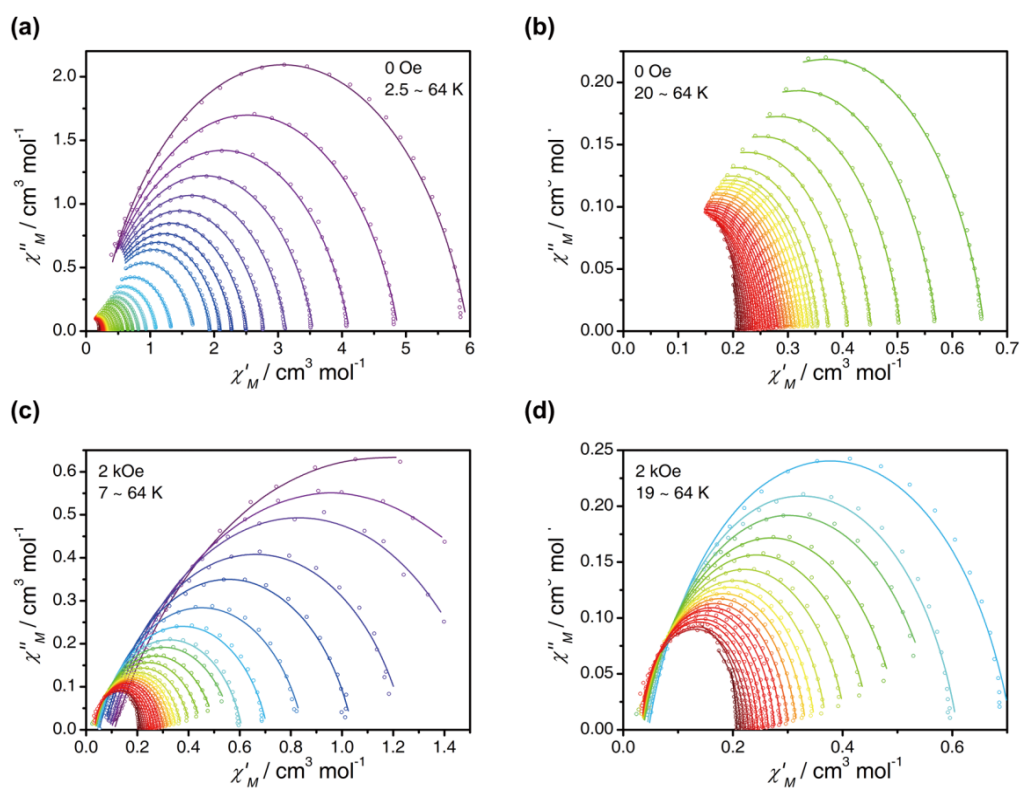

**Figure S7.** Cole-Cole plots for the ac susceptibilities under 0 and 2 kOe dc field for **1** (Left: full-temperature range; Right: high-temperature range). The solid lines are the best fit for the generalized Debye model.

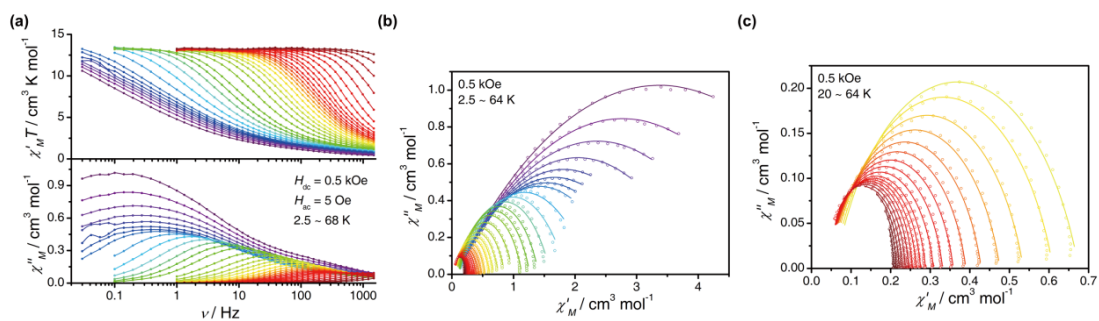

**Figure S8.** (a) The frequency dependence of the in-phase product ($\chi'_M T$) and out-of-phase ($\chi''_M$) for **1** at 0.5 kOe dc field with the temperature of 2.5−68 K. The solid lines are guides for the eyes. (b,c) Cole-Cole plots for the ac susceptibilities under 0.5 kOe dc field (b: full-temperature range; c: high-temperature range). The solid lines are the best fit for the generalized Debye model.

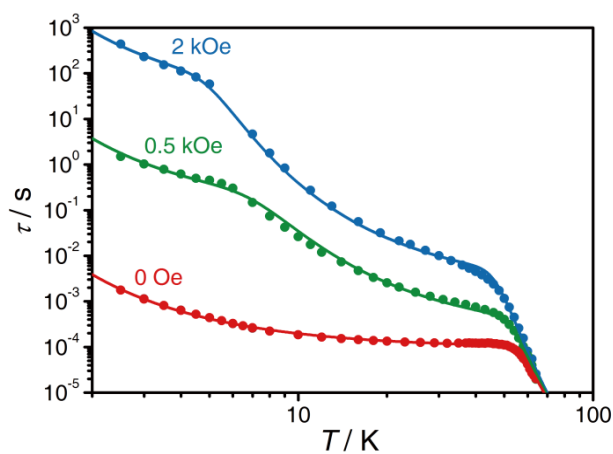

**Figure S9.** The temperature-dependent relaxation times (log-log scale plot of $\tau^{-1}$ vs. $T$) for **1** under indicated magnetic fields. The solid lines are fits with equation 3 for 0, 0.5 and 2 kOe dc field, respectively.

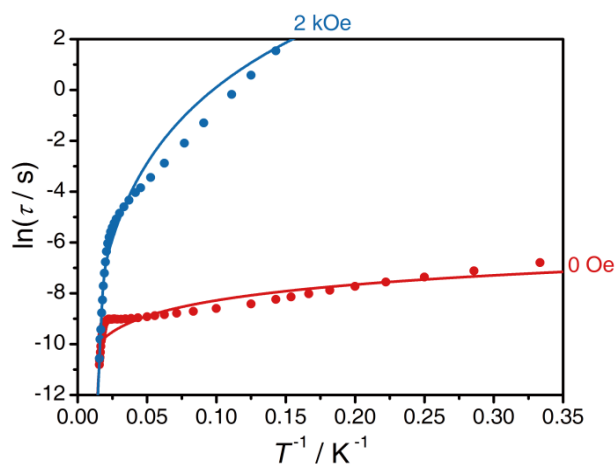

**Figure S10.** The temperature-dependent relaxation times for **1**. The data are fitted by a sum of Orbach + Raman processes, and show a serious deviation.

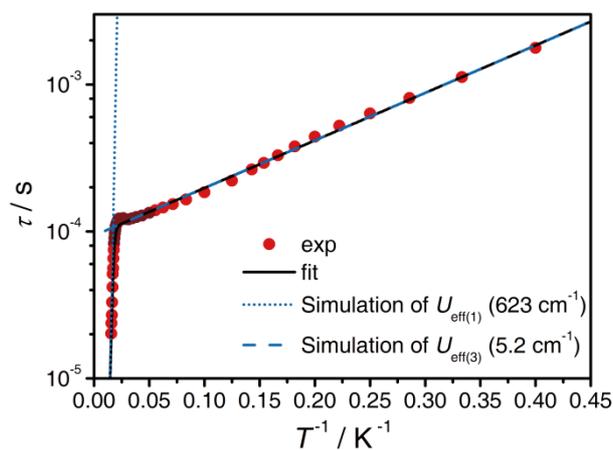

**Figure S11.** (a) The plots of relaxation times of **1** under zero dc field. The black solid line represents a sum of two Arrhenius law, giving $U_{eff(1)}$ = 623 cm$^{-1}$, $\tau_{0(1)}$ = 2.1 × 10$^{-11}$ s (blue dotted line) and $U_{eff(3)}$ = 5.2 cm$^{-1}$, $\tau_{0(3)}$ = 9.4 × 10$^{-5}$ s (blue dashed line). We are not currently certain on the origin of the bump highlighted in dark brown, and these data have been omitted from the fit in order to correctly estimate $U_{eff(1)}$ and $U_{eff(3)}$.

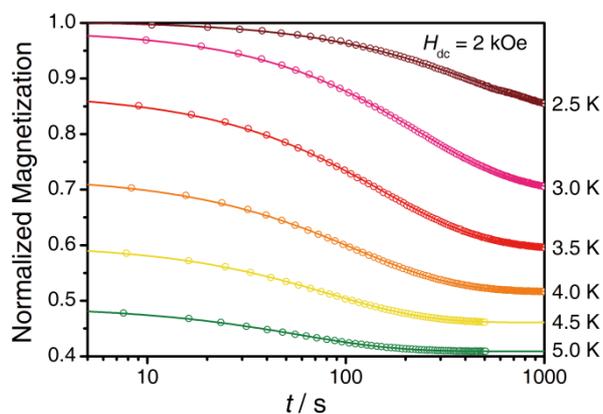

**Figure S12.** Various temperature magnetization decay and the best fit curves at 2 kOe.

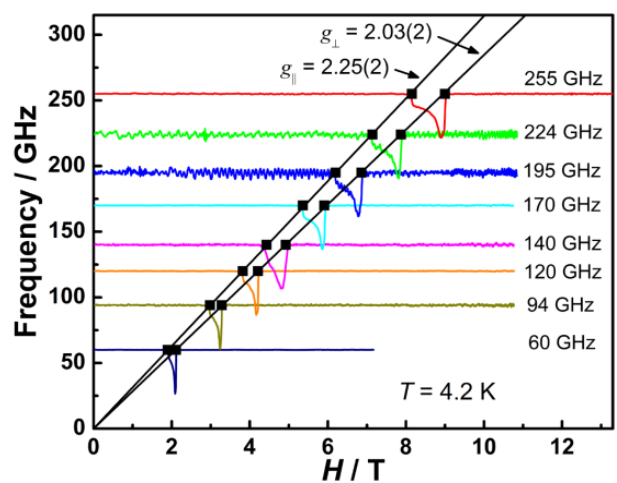

**Figure S13.** HF-EPR spectra of **2** at 4.2 K and various frequencies. The spectra are offset in proportion to the frequency. Black dots represent the resonance fields for each spectrum. Solid lines are the linear fit to each resonance branch. The obtained g-values are $g_{\parallel}$ = 2.25(2) and $g_{\perp}$ = 2.03(2).

## S4. Theoretical Calculations

***Broken-symmetry DFT calculations.***

To obtain an estimate of the magnetic coupling between the copper centres, we performed DFT calculations on the experimental crystal structure of YCu$_5$ with no optimisation in Gaussian09d,[4] using the well-known B3LYP[5] hybrid functional and the following basis sets: LANL2DZ[6] pseudopotential for yttrium, cc-pVTZ[7] for copper and cc-pVDZ[8] for all other atoms.

We assume the isotropic Heissenberg-Van Vleck-Dirac (HDVV) model spin Hamiltonian to be a good model, and proceed to map the expectation energy values of the different ferro- (FM) and antiferromagnetic (AFM) solutions to the exact non-relativistic, time-independent Hamiltonian. This is formally the same as employing the eigenstates of the simplified Ising model spin Hamiltonian. The molecular structure suggests that one may expect two different magnetic coupling constants within the Cu$^{II}_5$ pentagon, which can be represented by the following HDVV Hamiltonian, where $\langle i,j \rangle$ and $\langle k,l \rangle$ are the indexes of the *syn-syn* (3 pairs) and *syn-anti* (2 pairs) pairs of Cu ions, respectively, accounting for the orientation of the pyridine ligands.

$$\hat{H}^{HDVV} = -2 \sum_{\langle i,j \rangle} J\, \hat{S}_i \cdot \hat{S}_j - 2 \sum_{\langle k,l \rangle} J'\, \hat{S}_k \cdot \hat{S}_l$$

Making use of the set of functions presented in Table S6, one can obtain expressions for the HDVV expectation values in terms of the magnetic coupling constants (Table S7). Combining the calculated energy differences in Table S6 with the associated analytical expressions in Table S7, one can solve all set of linearly independent equations to obtain the different coupling constant values, as reported in Table S8. While there is some dispersion of $J$ and $J'$, our DFT calculations give $J = -98(4)$ cm$^{-1}$ and $J' = -66(9)$ cm$^{-1}$, supporting a two-J model for fitting the magnetic data. The ratio $J/J' = 1.50$ from DFT is consistent with that found experimentally of $J/J' = 1.55$.

**Table S6**. DFT computed energies and corresponding energy differences with respect to the ferromagnetic solution. The ordering of spins in the functions $|f\rangle$ is the same as in Scheme 1. Labels are used to indicate which energy differences have been used in Table S8.

| $\|f\rangle \equiv \|S_{Cu1}, S_{Cu2}, S_{Cu3}, S_{Cu4}, S_{Cu5}\rangle$ | Energy (a.u.) | $\Delta E$ (cm$^{-1}$) | Label |
|---|---|---|---|
| $\|\alpha\alpha\alpha\alpha\alpha\rangle$ | -13545.64838 | | |
| $\|\beta\alpha\alpha\alpha\alpha\rangle$ | -13545.64916 | 171.52 | 1 |
| $\|\alpha\beta\alpha\alpha\alpha\rangle$ | -13545.64928 | 199.09 | 2 |
| $\|\alpha\alpha\beta\alpha\alpha\rangle$ | -13545.64910 | 159.98 | 3 |
| $\|\alpha\beta\alpha\beta\alpha\rangle$ | -13545.64999 | 353.88 | 4 |
| $\|\alpha\beta\alpha\alpha\beta\rangle$ | -13545.64999 | 354.41 | 5 |
| $\|\beta\alpha\alpha\beta\alpha\rangle$ | -13545.64986 | 326.03 | 6 |

**Table S7**. Specification of FM and AFM HDVV expectation values and associated energy differences used to extract the two different magnetic coupling constants. Note that the ordering of spins in the functions $|f\rangle$ is the same as in Scheme 1, and the one to one correspondence with the entries in Table S6.

| $\|f\rangle$ | $\langle f\|\hat{H}^{HDVV}\|f\rangle$ | $\langle FM\|\hat{H}^{HDVV}\|FM\rangle - \langle AFM\|\hat{H}^{HDVV}\|AFM\rangle$ |
|---|---|---|
| $\|\alpha\alpha\alpha\alpha\alpha\rangle$ | $-1/2\,(2J + 3J')$ | |
| $\|\beta\alpha\alpha\alpha\alpha\rangle$ | $-1/2 \cdot J'$ | $-J - J'$ |
| $\|\alpha\beta\alpha\alpha\alpha\rangle$ | $-1/2 \cdot J'$ | $-J - J'$ |
| $\|\alpha\alpha\beta\alpha\alpha\rangle$ | $-1/2\,(2J - J')$ | $-2 \cdot J'$ |
| $\|\alpha\beta\alpha\beta\alpha\rangle$ | $1/2\,(2J + J')$ | |
| $\|\alpha\beta\alpha\alpha\beta\rangle$ | $1/2\,(2J + J')$ | $-2 \cdot (J + J')$ |
| $\|\beta\alpha\alpha\beta\alpha\rangle$ | $1/2\,(2J + J')$ | |

**Table S8**. Coupling constants extracted for the different pathways, using the different set of equations available, as labelled in Table S6.

| Labels of energy differences used | $J$ (cm$^{-1}$) | $J'$ (cm$^{-1}$) |
|:---:|:---:|:---:|
| 1 and 2 | -100 | -72 |
| 2 and 3 | -100 | -60 |
| 2 and 6 | -100 | -63 |
| 2 and 4 | -100 | -55 |
| 3 and 4 | -97 | -63 |
| 1 and 4 | -91 | -80 |

**Table S9**. Spin states and energies for {YCu$_5$}.

| $S_T$ | Energy (cm$^{-1}$) |
|:---:|:---:|
| 1/2 | 0 |
| 1/2 | 51 |
| 1/2 | 143 |
| 3/2 | 156 |
| 3/2 | 173 |
| 1/2 | 271 |
| 1/2 | 273 |
| 3/2 | 283 |
| 3/2 | 299 |
| 5/2 | 362 |

*CASSCF-SO calculations*

We employed Molcas 8.0[9] to perform CASSCF-SO calculations on the experimental crystal structure of {DyCu$_5$} where the Cu$^{II}$ ions were substituted by diamagnetic Zn$^{II}$ ions. Basis sets from the ANO-RCC library[10,11] were used with VTZP quality for the Dy atom, VDZP quality for the nearest oxygen atoms and VDZ for all other atoms. The two electron integrals were Cholesky decomposed with a threshold of 10$^{-8}$ to save disk space and computational resources. The state-averaged CASSCF orbitals of the 21 sextets were optimised with the RASSCF module. These orbitals were then used to obtain the 224 quartets by means of a configuration-interaction-only (no orbital optimisation). These two sets of spin-free states were then used to construct and diagonalise the spin-orbit coupling Hamiltonian in the basis of 21 and 128 sextets and quartets, respectively, with the RASSI module. The crystal field decomposition of the ground $J = {}^{15}/_2$ multiplet of the $^6$H term was performed with the SINGLE_ANISO module.[12]

**Table S10**. Crystal field splitting of the $^6$H$_{15/2}$ multiplet for Dy in DyCu$_5$, calculated with CASSCF-SO.

| Ab initio energy (cm$^{-1}$) | $g_x$ | $g_y$ | $g_z$ | $g_z$ Angle (°) | Crystal Field Energy (cm$^{-1}$) | Crystal Field Wavefunction |
|:---:|:---:|:---:|:---:|:---:|:---:|:---:|
| 0 | 0.00 | 0.00 | 19.89 | - | 0 | 100%$\|\pm 15/2\rangle$ |
| 314 | 0.00 | 0.00 | 17.10 | 1.3 | 313 | 100%$\|\pm 13/2\rangle$ |
| 511 | 1.43 | 7.35 | 13.46 | 89.9 | 510 | 89%$\|\pm 1/2\rangle$ + 5%$\|\mp 1/2\rangle$ + 4%$\|\mp 3/2\rangle$ + 2%$\|\pm 3/2\rangle$ |
| 541 | 0.19 | 0.49 | 13.61 | 4.3 | 542 | 92%$\|\pm 11/2\rangle$ + 6%$\|\pm 3/2\rangle$ + 2%$\|\pm 1/2\rangle$ |
| 556 | 1.86 | 4.07 | 4.62 | 21.6 | 556 | 65%$\|\pm 3/2\rangle$ + 22%$\|\mp 3/2\rangle$ + 4%$\|\pm 11/2\rangle$ + 3%$\|\pm 5/2\rangle$ + 2%$\|\pm 1/2\rangle$ + 1%$\|\mp 1/2\rangle$ + 1%$\|\mp 5/2\rangle$ |
| 621 | 0.72 | 1.82 | 6.82 | 6.3 | 622 | 91%$\|\pm 5/2\rangle$ + 5%$\|\pm 7/2\rangle$ + 3%$\|\pm 3/2\rangle$ + 1%$\|\pm 9/2\rangle$ |
| 657 | 0.23 | 0.39 | 12.84 | 25.8 | 656 | 88%$\|\pm 9/2\rangle$ + 11%$\|\pm 7/2\rangle$ |
| 677 | 0.21 | 0.39 | 13.42 | 45.9 | 676 | 84%$\|\pm 7/2\rangle$ + 11%$\|\pm 9/2\rangle$ + 5%$\|\pm 5/2\rangle$ |

**Table S11.** CASSCF-SO calculated CF parameters (cm$^{-1}$)

| k | q | B(k,q) |
|---|---|---|
| 2 | -2 | 0.94703622185356E-02 |
| 2 | -1 | 0.30252465348580E-01 |
| 2 | 0 | -0.30197369300000E+01 |
| 2 | 1 | -0.26997414970545E+00 |
| 2 | 2 | 0.15815788570198E+00 |
| 4 | -4 | -0.25058550502222E-02 |
| 4 | -3 | 0.43492341289488E-02 |
| 4 | -2 | 0.43348325188428E-03 |
| 4 | -1 | -0.15352173492993E-02 |
| 4 | 0 | -0.11725362007375E-01 |
| 4 | 1 | 0.30283268386897E-02 |
| 4 | 2 | -0.13493839249758E-02 |
| 4 | 3 | -0.84289430330379E-02 |
| 4 | 4 | 0.79918998501252E-03 |
| 6 | -6 | -0.76411284866855E-05 |
| 6 | -5 | -0.69517717456357E-04 |
| 6 | -4 | -0.10660970745767E-04 |
| 6 | -3 | 0.17406687745382E-04 |
| 6 | -2 | -0.43435319915386E-06 |
| 6 | -1 | 0.24789033577866E-06 |
| 6 | 0 | 0.27771888397897E-04 |
| 6 | 1 | -0.57161468366960E-04 |
| 6 | 2 | -0.47113012046145E-06 |
| 6 | 3 | -0.43314822885444E-04 |
| 6 | 4 | 0.79152086655417E-06 |
| 6 | 5 | -0.88315016262650E-04 |
| 6 | 6 | 0.78251082173440E-05 |

**Table S12.** Lowest lying 14 pseudo-doublets of {DyCu$_5$}.

| Energy (cm$^{-1}$) | $m_S$ (Cu$_5$) | $m_J$ (Dy) | $M_z$ |
|---|---|---|---|
| 0.0 | ±0.5 | ±7.5 | ±8 |
| 4.4 | ∓0.5 | ±7.5 | ±7 |
| 50.6 | ±0.5 | ±7.5 | ±8 |
| 55.0 | ∓0.5 | ±7.5 | ±7 |
| 142.9 | ±0.5 | ±7.5 | ±8 |
| 147.3 | ∓0.5 | ±7.5 | ±7 |
| 152.0 | ±1.5 | ±7.5 | ±9 |
| 156.3 | ±0.5 | ±7.5 | ±8 |
| 160.7 | ∓0.5 | ±7.5 | ±7 |
| 165.1 | ∓1.5 | ±7.5 | ±6 |
| 168.7 | ±1.5 | ±7.5 | ±9 |
| 173.1 | ±0.5 | ±7.5 | ±8 |
| 177.5 | ∓0.5 | ±7.5 | ±7 |
| 181.9 | ∓1.5 | ±7.5 | ±6 |